\documentclass[twocolumn]{aastex62}
\usepackage{graphicx}
\usepackage{amsmath}
\usepackage{amssymb}
\usepackage{color}
\usepackage{hyperref}

\gdef\xx[#1]{\textcolor{red}{#1}}
\def\code#1{\texttt{#1}}

\gdef\ma{mag\,arcsec$^{-2}$}

\newcommand{\GG}[1]{}

\lefthead{van Dokkum et al.}
\righthead{}
\submitjournal{PASP}
\begin{document}

\newcommand\XXX[1]{{\textcolor{red}{\textbf{x\ #1\ x}}}}

\title{\large \bf Multi-resolution filtering: an empirical method for isolating
faint, extended emission in Dragonfly data and other low resolution images}

\correspondingauthor{Pieter van Dokkum}

\author[0000-0002-8282-9888]{Pieter van Dokkum}
\affiliation{Astronomy Department, Yale University, 52 Hillhouse Ave,
New Haven, CT 06511, USA}

\author{Deborah Lokhorst}
\affiliation{Department of Astronomy \& Astrophysics,
University of Toronto, 50 St.\ George
Street, Toronto, ON M5S 3H4, Canada}

\author{Shany Danieli}
\affiliation{Physics Department, Yale University, 52 Hillhouse Ave,
New Haven, CT 06511, USA}

\author[0000-0001-9592-4190]{Jiaxuan Li}
\affiliation{Department of Astronomy, Peking University,
5 Yiheyuan Road, Haidian District, Beijing 100871, China}

\author{Allison Merritt}
\affiliation{Max-Planck-Institut f\"ur Astronomie,
K\"onigstuhl 17, D-69117 Heidelberg, Germany}

\author{Roberto Abraham}
\affiliation{Department of Astronomy \& Astrophysics,
University of Toronto, 50 St.\ George
Street, Toronto, ON M5S 3H4, Canada}

\author{Colleen Gilhuly}
\affiliation{Department of Astronomy \& Astrophysics,
University of Toronto, 50 St.\ George
Street, Toronto, ON M5S 3H4, Canada}

\author[0000-0003-4970-2874]{Johnny P. Greco}
\altaffiliation{NSF Astronomy \& Astrophysics Postdoctoral Fellow}
\affiliation{Center for Cosmology and AstroParticle Physics (CCAPP), The Ohio State University, Columbus, OH 43210, USA}

\author{Qing Liu}
\affiliation{Department of Astronomy \& Astrophysics,
University of Toronto, 50 St.\ George
Street, Toronto, ON M5S 3H4, Canada}

\begin{abstract}

We describe an empirical, self-contained
method to isolate  faint, large-scale emission in imaging data
of low spatial resolution.
Multi-resolution filtering (MRF) uses independent data of superior
spatial resolution and point source depth to create a model for all compact and
high surface brightness objects in the field. This model is convolved with an appropriate kernel and
subtracted from the low resolution image. The halos of bright stars are
removed in a separate step and artifacts are masked. The resulting
image only contains extended emission fainter than a pre-defined surface brightness limit.
The method was developed for the Dragonfly Telephoto Array, which produces images
that have excellent
low surface brightness sensitivity but poor spatial resolution. 
We demonstrate
the MRF technique using
Dragonfly images of a satellite of the spiral galaxy M101,
the tidal debris surrounding M51, 
two ultra-diffuse galaxies in the Coma cluster, and the galaxy NGC\,5907.
As part of the analysis we present a newly-identified
very faint galaxy in the filtered Dragonfly image of the M101 field.
We also discuss variations of the technique for cases when no
low resolution data are available (self-MRF and cross-MRF).
The method is implemented in  \code{mrf}, an open-source MIT licensed Python
package.\footnote{\url{https://github.com/AstroJacobLi/mrf}}

\end{abstract}

\section{Introduction}

Phenomena observed at low surface brightness
hold the potential to inform many areas of astronomy.
Examples include diffuse cirrus emission, which
provides information on the interstellar medium in the Milky Way
({Miville-Desch{\^e}nes} {et~al.} 2016);
light echoes, which can be used to characterize ancient
supernovae ({Rest} {et~al.} 2005, 2008); and the dust and gas associated with
solar system bodies such as comets
(e.g., {Sekanina} \& {Miller} 1976; {Pittichov{\'a}} {et~al.} 2008).

Perhaps the richest returns have come from extragalactic studies.
The faintest dwarf galaxies generally have extremely low surface brightness, well
beyond the limits of conventional blank-sky surveys (see {McConnachie} 2012). Accordingly, so far most of them
have been discovered through the detection of their individual giant stars
(e.g., {Belokurov} {et~al.} 2007; {Bechtol} {et~al.} 2015; {M{\"u}ller}
et al.\ 2019), something that
is only possible if they are relatively nearby ($D\lesssim 5$\,Mpc; {Danieli}, {van Dokkum}, \&  {Conroy} 2018). 
Furthermore, from the 1980s onward it has become apparent that many relatively
luminous galaxies were missing from traditional surveys due to their low surface brightness. These include
classical low surface brightness galaxies (LSBs), which are typically gas-rich spiral
galaxies with faint, large disks (e.g., {van der Hulst} {et~al.} 1993; {de Blok} {et~al.} 2001; {Schombert}, {McGaugh}, \&  {Maciel} 2013),
as well as large spheroidal objects with little or no gas ({Impey}, {Bothun}, \& {Malin} 1988; {Bothun}, {Impey}, \& {Malin} 1991).
These ``ultra diffuse galaxies'' (UDGs)
turn out to be surprisingly common ({van Dokkum} {et~al.} 2015; {van der Burg} {et~al.} 2017; {Danieli} \& {van Dokkum} 2019), and
display a bewildering variety of properties (e.g., {Merritt} {et~al.} 2016b; {Beasley} {et~al.} 2016; {van Dokkum} {et~al.} 2018).
Other examples of extragalactic low surface brightness regimes are studies of stellar halos
({Zibetti}, {White}, \&  {Brinkmann} 2004; {de Jong} 2008; {Tal} \& {van Dokkum} 2011; {Duc} {et~al.} 2015; {Merritt} {et~al.} 2016a; {Trujillo} \& {Fliri} 2016), tidal features ({Arp} 1966; {Malin} \& {Hadley} 1997; {van Dokkum} 2005; {Mart{\'{\i}}nez-Delgado} {et~al.} 2010; {Atkinson}, {Abraham}, \&  {Ferguson} 2013), and intracluster light
({Gonzalez} {et~al.} 2000; {Zibetti} {et~al.} 2005; {Mihos} {et~al.} 2005; {DeMaio} {et~al.} 2015; {Montes} \& {Trujillo} 2018).


Efforts to detect and characterize low surface brightness emission have mostly
made use of the
excellent image quality of mirror telescopes
equipped with wide-field  CCD cameras. Low surface brightness galaxies and structures
have been identified in data from, among others, the 
Burrell Schmidt telescope (Mihos et al.\ 2005; Watkins, Mihos, \& Harding 2015), the
Sloan Digital Sky Survey ({Bakos} \& {Trujillo} 2012; {Fliri} \& {Trujillo} 2016),
the Canada France Hawaii Telescope ({Duc} {et~al.} 2014, 2015), and the Subaru Telescope
({Koda} {et~al.} 2015; {Mowla} {et~al.} 2017; {Greco} {et~al.} 2018a). An
advantage of using such relatively high resolution imaging data sets is that
the contrast between low surface brightness objects and everything else
in the field is maximized. As shown by, e.g., {Greco} {et~al.} (2018b) the faint glow
of low surface brightness galaxies can be isolated
by applying a variation of low pass filtering.
Furthermore, as there are many applications of such data
low surface brightness surveys often piggy-back onto other programs or use
publicly available archival data.

A disadvantage is that the
mosaiced cameras that are used in these surveys
are comprised of many individual detectors, and it is difficult
to achieve accurate flat fielding and sky subtraction on scales that exceed the
size of an individual chip (see, e.g., Fig.\ 4 in {Aihara} {et~al.} 2019).
Additionally, the mirrors, open structure,
and complex light path of modern reflectors can lead to artifacts and produce
point spread functions
(PSFs) with complex wings containing significant power.
As shown in {Slater} {et~al.} (2009) scattered light from field stars and the centers of galaxies
produces a low surface brightness floor of $\approx 29.5$\,mag\,arcsec$^{-2}$ even when
all other effects are controlled for.
These issues can be mitigated to an impressive degree thanks to innovative observing
and reduction strategies. Examples of projects that reach depths well beyond the
traditional limits of CCD imaging are the Burrell Schmidt Virgo Cluster Survey
(Mihos et al.\ 2017), the MATLAS survey with
CFHT/MegaCAM (Duc et al.\ 2015; Karabal et al.\ 2017),
the Fornax Deep Survey (Iodice et al.\ 2015, 2019), and the IAC Stripe 82 Legacy
Project (Fliri \& Trujillo 2016).

An alternative approach\footnote{A hybrid approach is to harnass small reflecting telescopes for low surface brightness studies. Examples are the HERON 0.7\,m telescope (Rich et al.\ 2019) and the 0.5\,m telescopes used by advanced amateurs (see Martinez-Delgado et al.\ 2010).} is to detect low surface brightness emission in low resolution
images, such as those delivered by the Dragonfly Telephoto Array 
({Abraham} \& {van Dokkum} 2014) and the Huntsman telescope (Spitler et al.\ 2019).\footnote{An early
implementation of this idea was the ``Parking Lot Camera'', which obtained deep images of the
Magellanic Clouds with
a pixel size of $73\arcsec$ (Bothun \& Thompson 1988).}
In these data the PSF is matched to the structures
of interest,
which maximizes the detection signal-to-noise (S/N) ratio without the need to
apply a low pass filter
(see {Irwin} 1985). Specifically, Dragonfly uses monolithic detectors
with $2\farcs 8$ native pixels
and a field of view of $3^{\arcdeg}\times 2^{\arcdeg}$, and with 48 independent sight lines
flat fielding and sky modeling are well-controlled on scales $\lesssim 45\arcmin$. 
Furthermore, the Canon 400\,mm f/2.8 IS II telephoto lenses that constitute the heart
of Dragonfly are excellent for low surface brightness imaging: the light path
is fully enclosed and thanks to the all-refractive design and ``sub-wavelength''
anti-reflection coatings the PSF is well-behaved with low power in the wings
(see {Abraham} \& {van Dokkum} 2014).

The downside of the Dragonfly approach is {\em blending}: owing to the typical
FWHM of $\sim 5\arcsec$
stars and compact galaxies take up a much larger fraction of the detector than
would be the case for seeing-limited data.\footnote{The delivered FWHM is 
currently limited by the large native pixel size of the Dragonfly cameras; the lenses are
able to deliver sharper images (see Abraham \& van Dokkum 2014).} Furthermore, groups of faint stars and galaxies
masquerade as spatially-extended low surface brightness galaxies. As an example,
in {van Dokkum} {et~al.} (2015)
the final list of 47 UDGs was extracted from a parent sample of 6624 faint,
spatially-extended Dragonfly detections.

Here we present a method that combines the advantages of seeing-limited (``high resolution'')
observations with those of low resolution Dragonfly-like data. The high resolution data
are used to create a model for all the emission that is {\em not} low surface brightness.
This model is then convolved with a suitable kernel to match the resolution of the Dragonfly
data and subtracted. All faint galaxies, stars, and blends are removed in this residual
image and genuine low surface brightness features can be identified and quantified.
In the following sections we discuss the general method and then the
specific implementation in the Python package {\tt mrf}. We stress that this paper does not
discuss how to {\em detect} the remaining low surface brightness
emission, or how to quantify the uncertainties in the detected flux. The outputs of
the {\tt mrf} code include
the flux that was subtracted for each pixel, and we encourage researchers to
take this subtracted flux properly into account when assessing the residual images in
the context of specific science projects.

\section{Methodology}

\subsection{Multi-resolution filtering}
\label{mrf_method.sec}

The method is closely related to other image matching algorithms, such as those employed
for transient detection (e.g., {Alard} \& {Lupton} 1998; {Miller}, {Pennypacker}, \&  {White} 2008; {Zackay}, {Ofek}, \& {Gal-Yam} 2016), photometry in extragalactic
survey fields (see, e.g., {Labb{\' e}} {et~al.} 2003; {Merlin} {et~al.} 2016), and ``forced'' photometry of SDSS sources in WISE data ({Lang}, {Hogg}, \& {Schlegel} 2016). All these techniques perform some kind of PSF matching between datasets, with one PSF-matched dataset serving as a point of reference{} for the other.
In our method
the detection of low surface brightness emission is not performed directly in the low
resolution image $I^{\rm L}$ but in a residual image $R$:
\begin{equation}
R = I^{\rm L} - F^{\rm L},
\end{equation}
with $F^{\rm L}$ a flux model that has the same spatial resolution as $I^{\rm L}$. This low resolution
model is created by convolving a high resolution model with a kernel:
\begin{equation}
F^{\rm L} = F^{\rm H} \ast K.
\end{equation}
The high resolution model $F^{\rm H}$ is based on a high resolution image $I^{\rm H}$ and is composed of stars, unresolved or marginally-resolved galaxies, and the high surface brightness regions of resolved galaxies.\footnote{The point-source depth of the high resolution image should exceed that of the low resolution image. For Dragonfly data, this criterion is generally easily met as the $5\sigma$ point source depth of typical Dragonfly data is only $g\sim 23$ (e.g., Merritt et al.\ 2016a).}
This model can be constructed in a variety of ways. The simplest is to isolate all pixels in
$I^{\rm H}$ above a per-pixel brightness limit.
However, this method is susceptible to mis-identifying noise peaks as objects, and it ignores
the fact that $I^{\rm H}$ is typically a PSF-convolved image itself with a finite spatial resolution (characterized by a point spread function
$P^{\rm H}$).
The contrast between compact sources and the rest of the image is maximized by
convolving the image with a filter that is a mirrored version of $P^{\rm H}$
(with the $x$ and $y$ coordinates swapped; see {Irwin} 1985). Next, objects can be
detected in this convolved image by finding groups of $\geq N$ connected pixels above a particular per-pixel brightness (see, e.g., {Lutz} 1979).
In practice, these steps are incorporated in the Source Extractor
(SExtractor; {Bertin} \& {Arnouts} 1996) software, and
as shown in \S\,\ref{implement.sec} $F^{\rm H}$ can be created with the help of the segmentation map and catalog produced by this program.

Likewise, the kernel $K$ can be generated in multiple ways. {Alard} \& {Lupton} (1998) show that a least-squares method applied to all pixels in the image can produce the optimal convolution kernel, as long as the kernel can be approximated by Gaussian decomposition. Another robust method is to forego solving for $K$ and perform a double convolution: $R = I^{\rm L} \ast P^{\rm H} - F^{\rm H} \ast P^{\rm L}$
(e.g., {Gal-Yam} {et~al.} 2008). This method is, in principle, well-suited to our situation as the precise
form of $P^{\rm H}$ is unimportant
and can be approximated by a simple Moffat or Gaussian function. However, in practice the Dragonfly
PSF $P^{\rm L}$ is dominated by focusing and guiding errors on the relevant spatial scales, which means
that this approach still requires an independent method to determine the PSF for each frame.
In light of these considerations we adopt an extension of 
the ``classical'' approach to generating $K$:
\begin{equation}
\label{fft1.eq}
K = f \left(\frac{\mathcal F(P^{\rm L})}{\mathcal F(P^{\rm H})}\right) \approx
f\left(\frac{\mathcal F(I^{\rm L})}{\mathcal F(I^{\rm H})}\right),
\end{equation}
with $\mathcal F$ denoting the forward Fourier transform and $f$ the inverse Fourier transform
(see, e.g., {Phillips} \& {Davis} 1995).  This method is fast and easily implemented but 
has several downsides. It requires a window function to dampen high frequency noise
in the kernel, it can produce artifacts when large parts of the image are relatively
empty, and it is numerically unstable in the presence of noise.
In our case the results are not very sensitive to the choice of window
function as nearly all the power in the Dragonfly PSF is on scales of
$\lesssim 5$ native pixels (see {Abraham} \& {van Dokkum} 2014): the kernel does not need
to capture diffraction patterns or other complex structures as they
have very little power.\footnote{These complex structures
are only relevant for the modeling of bright, generally saturated,
stars; this is done in a separate step (see \S\,\ref{implement.sec}).}
The other two issues are mitigated by
generating
a set of $k$ kernels from small
image cutouts centered on relatively isolated, non-saturated bright stars, and
then letting $K$ be the median or mean of this set:
\begin{equation}
\label{fft2.eq}
K \approx \frac{1}{k}\sum_{i=1}^{i=k} f \left(\frac{\mathcal
F(I_i^{\rm L})}{\mathcal F(I_i^{\rm H})}\right),
\end{equation}
with $I_i$ the individual cutouts and typical values of $k=20-30$.
As shown in \S\,\ref{implement.sec} this method produces satisfactory results
in an automated way.

\begin{figure*}[htbp]
  \begin{center}
  \includegraphics[width=1.0\linewidth]{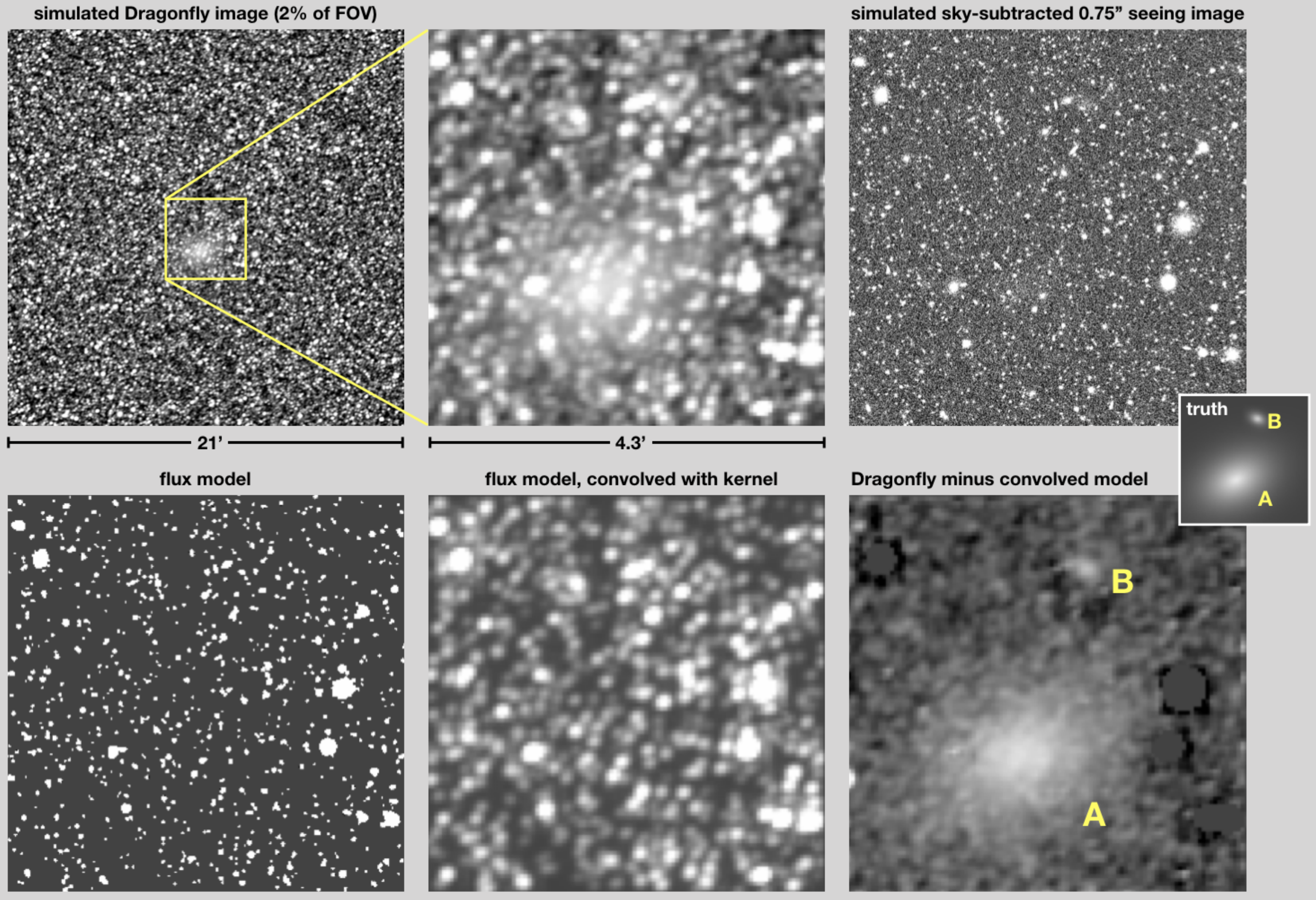}
  \end{center}
\vspace{-0.2cm}
    \caption{
Demonstration of multi-resolution filtering (MRF), using simulated data. {\em Top left and middle:} 
Small section of a simulated deep Dragonfly image containing two low surface brightness objects.
The image is dominated by stars and bright galaxies. {\em Top right:} Higher resolution image
with $0\farcs 75$ seeing. The sky was modeled and subtracted using standard SExtractor
settings. {\em Bottom left:} Flux model, consisting of all objects detected with
SExtractor in the high resolution image. {\em Bottom middle:} Flux model convolved with a kernel
determined from stars in the image. {\em Bottom right:} Residual Dragonfly image after
subtracting the convolved flux model. The two simulated galaxies are clearly visible and
can easily be detected in this image.
}
\label{mrf_sim.fig}
\end{figure*}

\subsection{Illustration using artificial data}

We illustrate the method with artificial data. An image of $21\farcm 3 \times 21 \farcm 3$ is
created, containing $10^4$ stars, $5\times 10^4$ galaxies, and a constant sky background.
The magnitude distributions
follow power laws, with slopes $0.6$ for the stars (appropriate for an isotropic
distribution) and $0.45$ for the galaxies (e.g., Lilly, Cowie, \& Gardner 1991). The 
distribution of half-light radii of the galaxies is uniform between 0 and $2\farcs 5$;
40\,\% follow $r^{1/4}$ (de Vaucouleurs) profiles and 60\,\% follow exponential profiles.
Two low surface brightness galaxies are added to the image. Their structure
and surface brightness profiles are modeled on that of the
UDG Dragonfly 44 ($R_{\rm e}=4.7$\,kpc, $\mu_{0,V}=24.1$\,mag\,arcsec$^{-2}$, and
Sersic index $n=0.94$; {van Dokkum} {et~al.} 2017), with one of the galaxies
placed at 20\,Mpc (galaxy A) and the other at 100\,Mpc (galaxy B).

The image
is generated at two resolutions and pixel sizes. The high resolution image has a pixel
size of $0\farcs 25$ and a PSF that follows a Moffat profile with $\beta=2.5$
and FWHM\,=\,$0\farcs 75$, typical
for wide field survey data of telescopes such as CFHT and Subaru. The
low resolution image is tailored to Dragonfly, with a pixel size of $2\farcs 5$ and
FWHM\,=\,$5\farcs 9$. The PSF is given an ellipticity of $\epsilon=0.2$ (and a position angle
of 30$^{\arcdeg}$) to simulate
guiding or focus errors.\footnote{This is representative of
early Dragonfly data; in the past few years we have improved our guiding
and focus procedures and currently Dragonfly produces PSFs with $\epsilon \lesssim 0.05$.}
To simulate realistic sky subtraction artifacts in the high resolution image
we modeled and subtracted its sky
using the default SExtractor settings, with a mesh size of $64\times 64$ pixels.

The low resolution image is shown in the top left panel of Fig.\ \ref{mrf_sim.fig}, with
a zoom of the area containing the low surface brightness galaxies in the top middle panel.
The images of these galaxies are highly contaminated by stars and galaxies. Galaxy A is visible
but it is difficult to discern its structure and size. Galaxy B cannot be identified reliably
in these data, as its appearance is similar to many other structures (blended faint
stars and galaxies) in the image. We quantified this by calculating the number of pixels
that are brighter than $\Sigma_{10}$, the mean galaxy surface brightness within
$10\times 10$\,pix$^2$.
Nine percent of all pixels in the image are brighter than $\Sigma_{10}$ of galaxy
A, and 36\,\% are brighter than $\Sigma_{10}$ of B.

The sky-subtracted high resolution image is shown at top right. Light from
stars and compact galaxies 
cover a much smaller fraction of the pixels owing to the $10\times$ narrower PSF. 
Galaxy B is now just-visible as a low surface brightness patch in the top part of the
frame. Galaxy A is also visible but is fainter and smaller than it should be due to the
sky subtraction. Also, the contrast between both galaxies and the background is low, and they
could easily be mistaken for ghosts, flat fielding errors, or other issues in the data
reduction.

The bottom panels show the multi-resolution filtering process applied to these data.
We performed the analysis with the {\tt mrf} code, which is described in detail
in the next Section. First a flux model is created by multiplying the high resolution
image by an object mask created with SExtractor (bottom left panel in Fig.\ \ref{mrf_sim.fig}).
Next, the flux model is convolved with a kernel to match the low resolution image
(bottom middle panel). The convolved model looks very similar to the low resolution image shown
above it, except that it does not include the low surface brightness objects.
The final step is the subtraction of this convolved model from the low resolution
image. The two low surface brightness objects are prominent in the
residual image, shown in the bottom right panel. The total fluxes of objects A and
B are recovered
to $\approx 95$\,\% and $\approx 80$\,\% respectively.

\begin{figure*}[htbp]
  \begin{center}
  \includegraphics[width=1.0\linewidth]{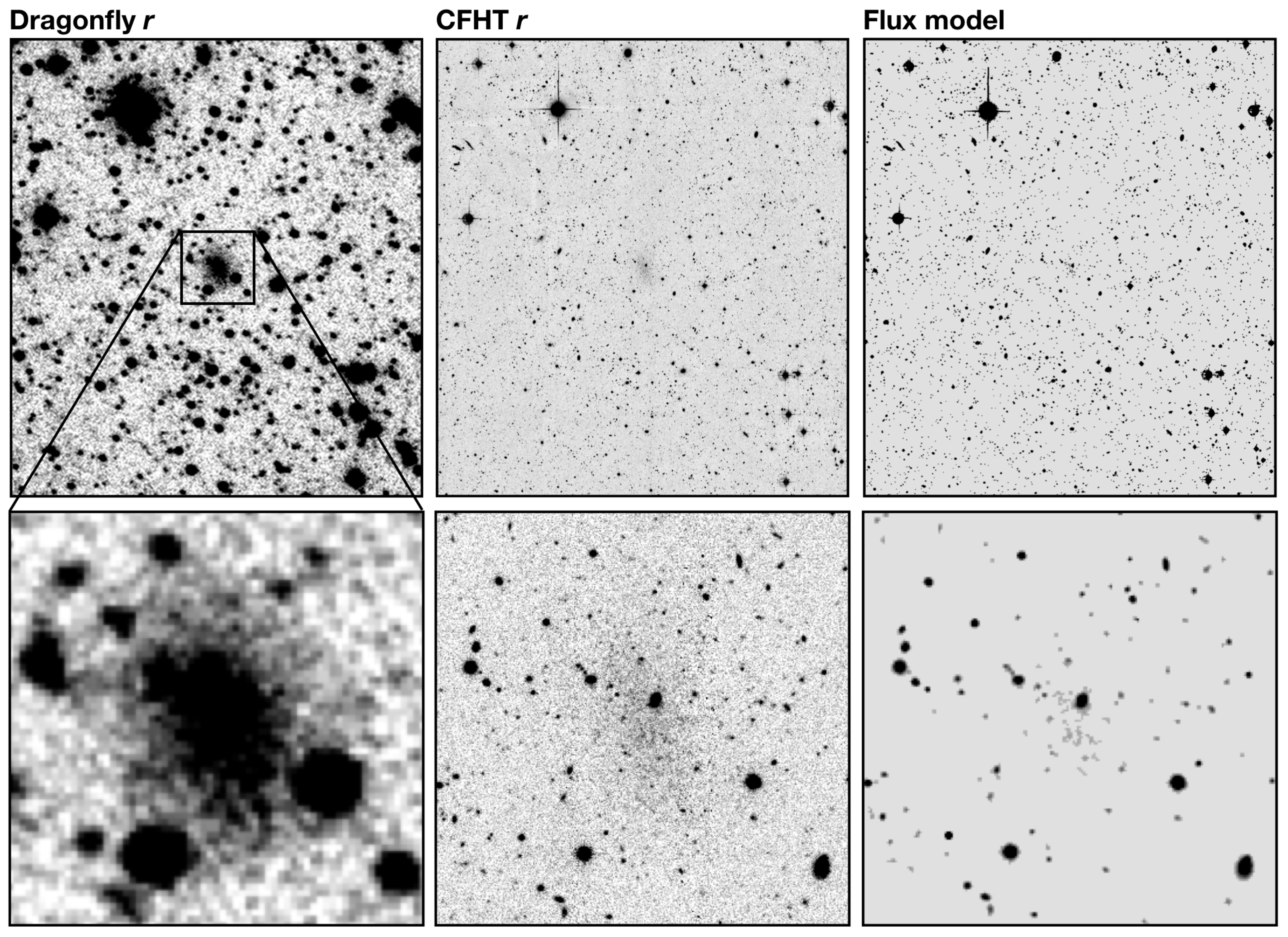}
  \end{center}
\vspace{-0.2cm}
    \caption{
{\em Top left:} Small section ($6\farcm 5 \times 7\farcm 3$) of the 10\,deg$^2$
Dragonfly $r$-band image of M101, centered on the satellite galaxy M101-DF3
({Merritt}, {van Dokkum}, \&  {Abraham} 2014; {Danieli} {et~al.} 2017). This image is referred to as $I^{\rm L}$ in
the text. {\em Top middle:} The same area as observed with CFHT in the context
of the CFHT Lensing Survey ($I^{\rm H}$). The resolution is far superior 
to that of Dragonfly, but M101-DF3 is fainter due to sky subtraction errors
on large scales. There is faint large scale PSF structure around the bright star
at the top of the frame. {\em Top right:} Initial flux model $F^{\rm H(3)}$,
generated by multiplying a SExtractor object mask by $I^{\rm H(3)}$. This model
contains all compact sources in the field and also some low surface brightness
emission from M101-DF3. {\em Bottom row:} Zoomed views of M101-DF3 ($2\farcm 1
\times 2\farcm 1$).
}
\label{df3_intro.fig}
\end{figure*}

\section{Implementation}
\label{implement.sec}

The method is implemented in the {\tt python} package
{\tt mrf}.\footnote{\url{https://github.com/AstroJacobLi/mrf}} The individual steps
are demonstrated using Dragonfly imaging of a low surface brightness
satellite of the nearby face-on
spiral galaxy M101. M101 is at a distance of 7\,Mpc ({Lee} \& {Jang} 2012), and it is one
of the nearest massive galaxies. It was observed in May and June 2013 in the
SDSS $g$ and $r$ filters with an eight-lens
configuration.
The pixel size of the reduced images is $2\farcs 0$, and the FWHM image quality is
$\approx 6\farcs 7$. The image is publicly available.\footnote{See \url{http://dragonflytelescope.org}.
We note that Dragonfly currently has 48 lenses. Also, our focusing and
guiding procedures have steadily improved over the past seven years,
and the delivered image quality is now typically $\approx 5\arcsec$.} 
In {van Dokkum}, {Abraham}, \&  {Merritt} (2014) and {Merritt} {et~al.} (2016a)
we measured the light profile of the galaxy,
constraining the mass of its stellar halo. In {Merritt} {et~al.} (2014) we presented
seven previously-uncataloged low surface brightness objects in the M101 field.
In follow-up studies with the {\it Hubble Space Telescope} ({\it HST}) we
found that four of these seven objects are intrinsically-large background galaxies
({Merritt} {et~al.} 2016b) and three are satellites of M101 ({Danieli} {et~al.} 2017).
Here we focus on a small region centered on one of these confirmed satellites,
M101-DF3 (see Fig.\ \ref{df3_intro.fig}).
The galaxy has an effective surface brightness of $\mu_{e,g}=27.4\pm 0.2$ and
an effective radius $r_e = 30\arcsec \pm 3\arcsec$ ({Merritt} {et~al.} 2014). Its
distance, as measured from the tip of the red giant branch, is $D=6.5 \pm 0.3$\,Mpc
({Danieli} {et~al.} 2017).

\subsection{High resolution M101 images}

The M101 field has publicly available imaging from the Canada France Hawaii
Telescope (CFHT), obtained in the context of the CFHT Lensing
Survey ({Heymans} {et~al.} 2012). The exposure times in $g$ and $r$
were 2500\,s per filter. The
CFHT images of M101-DF3 are shown in the middle panels of Fig.\ \ref{df3_intro.fig}.

The CFHT and Dragonfly data are brought to a common reference frame. This
frame has a finer pixel scale than the Dragonfly data, so that
the subpixel flux distribution of compact objects is properly modeled.
The low resolution
Dragonfly image $I^{\rm L}$ is sampled onto a grid with $1/3$ the pixel size
($0\farcs 67$ in the case of M101) using a third order polynomial for the interpolation.
This subsampled image is denoted $I^{\rm L(3)}$ and its size is $3n\times 3m$
if the size of $I^{\rm L}$ is $n\times m$.
For each filter the CFHT image
$I^{\rm H}$ is projected onto the $I^{\rm L(3)}$ frame using the WCS information
from the headers of both images and a third order polynomial for interpolation.
Prior to transforming $I^{\rm H}$ to $I^{\rm H(3)}$ the
image was binned $2\times 2$
and convolved with a $\sigma=1$\,pixel Gaussian, as this mitigates
projection errors going from the $0\farcs 186$ CFHT pixel
scale to that of $I^{\rm L(3)}$. In general, care should be taken that the pixel
size of the subsampled low resolution data is
within a factor of a few of the resolution of the high resolution data.

Next, the filter systems are matched. For both Dragonfly and CFHT
data are available in the $g$ and $r$ filters. However, even though the Dragonfly filter
curves are nearly identical to those used in SDSS and many other surveys, the
total system response is not. The SBIG STF-8300 cameras have a low quantum
efficiency in the blue, which means that the effective
Dragonfly $g$ is slightly redder than SDSS $g$. It is therefore necessary
to bring the CFHT images onto the same filter system as the Dragonfly data.

The {\tt mrf} code can interpolate between high resolution images in two filters.
It is assumed that
\begin{equation}
\label{color.eq}
I^{L(3)}(g_{\rm DF}) \propto I^{H(3)}(g_{\rm CFHT}) \left( \frac{ I^{H(3)}(g_{\rm CFHT}) }{ I^{H(3)}(r_{\rm CFHT}) } \right)^{\alpha_g},
\end{equation}
with an equivalent expression for the $r_{\rm DF}$ band.
The value of
$\alpha$ is determined from
synthetic photometry of stars in the Gunn \& Stryker (1983) atlas;
we find $\alpha_g \approx +0.05$  and $\alpha_r \approx +0.01$.

The ratio $I^{\rm H(3)}(g_{\rm CFHT})/I^{\rm H(3)}(r_{\rm CFHT})$ is only well-determined in
regions with a high S/N ratio in both filters, and a direct application
of Eq.\ \ref{color.eq} to the high resolution images
would lead to an extremely noisy interpolated image. We therefore create
a color correction image $C$ such that $I^{\rm H(3)}(g_{\rm DF}) =
I^{\rm H(3)}(g_{\rm CFHT}) \times
C^{\alpha}$. The correction image
is based on the SExtractor segmentation map (see below), with all pixels
belonging to an object set to the ratio of the SExtractor catalog fluxes in
$g_{\rm CFHT}$ and $r_{\rm CFHT}$. As
the catalog fluxes are much better determined than the flux in
an individual pixel this interpolation method is more robust than
a direct division. We caution that this procedure works best if the effective
wavelengths of the 
high resolution and low resolution data are reasonably close.

\subsection{Construction of the high resolution flux model}

Objects are identified in $I^{\rm H(3)}$ using
{\texttt{sep}} (Barbary 2016), which is a Python version of SExtractor.
The SExtractor parameters
can be adjusted for the particular dataset that is being analyzed;
as the aim is to detect compact
sources the code is typically run with a low minimum number of connected pixels (2 in
this example), a low threshold for deblending ($5\times 10^{-4}$), and
a fairly fine grid for the background model ($64\times 64$ pixels). The code
produces a catalog of objects with positions and fluxes, as well as a segmentation
map $S^{\rm H(3)}$. The value of each pixel in $S^{\rm H(3)}$ is that of the ID
of the object that the pixel is a part of. We create an object mask from $S^{\rm H(3)}$:
\begin{equation}
M^{\rm H(3)} = \begin{cases} 0, & \text{if}\ S^{\rm H(3)}=0 \\
1, & \text{otherwise.}
\end{cases}
\end{equation}

Next, objects that should {\em not} be subtracted
from the Dragonfly image are removed from the mask. An example where this
might be the case is a study of the stellar halos or tidal
features around a set of galaxies in the image; it is
then desirable to retain the high
surface brightness regions of these galaxies in the Dragonfly data.
The {\tt mrf} code can be supplied with 
objects that need to be retained, as specified by their
{\texttt{RA}}, {\texttt {DEC}} positions.
For each object
the ID is found in $S^{\rm H(3)}$ from the value of the pixel
that is closest to this
position. All pixels 
that have this same value in $S^{\rm H(3)}$
are identified, and set to zero in $M^{\rm H(3)}$.
Note that bright stars and low surface brightness objects are also removed
from the model, at a later stage (see \S\,3.4, 3.5, and 3.7).

The high resolution flux model is then
created by multiplying the image by the object mask:
\begin{equation}
F^{\rm H(3)} = I^{\rm H(3)} \times M^{\rm H(3)}.
\end{equation}
This initial flux model is shown in the rightmost panels
of Fig.\ \ref{df3_intro.fig}, for the
$r$ band.
The model contains stars, high surface brightness galaxies, and (parts of)
low surface brightness galaxies and features. It may contain artifacts such
as diffraction spikes, to the extent that SExtractor identifies them as
objects. It does not include any emission outside of the segmentation map,
that is, it ignores object flux beyond
the scaled Kron radius (see {Bertin} \& {Arnouts} 1996).
To account for this the mask can optionally be expanded by convolving
it with a circular
broadening function, prior to creating $F^{\rm H(3)}$.
This was not done in this example.

\begin{figure*}[htbp]
  \begin{center}
  \includegraphics[width=0.95\linewidth]{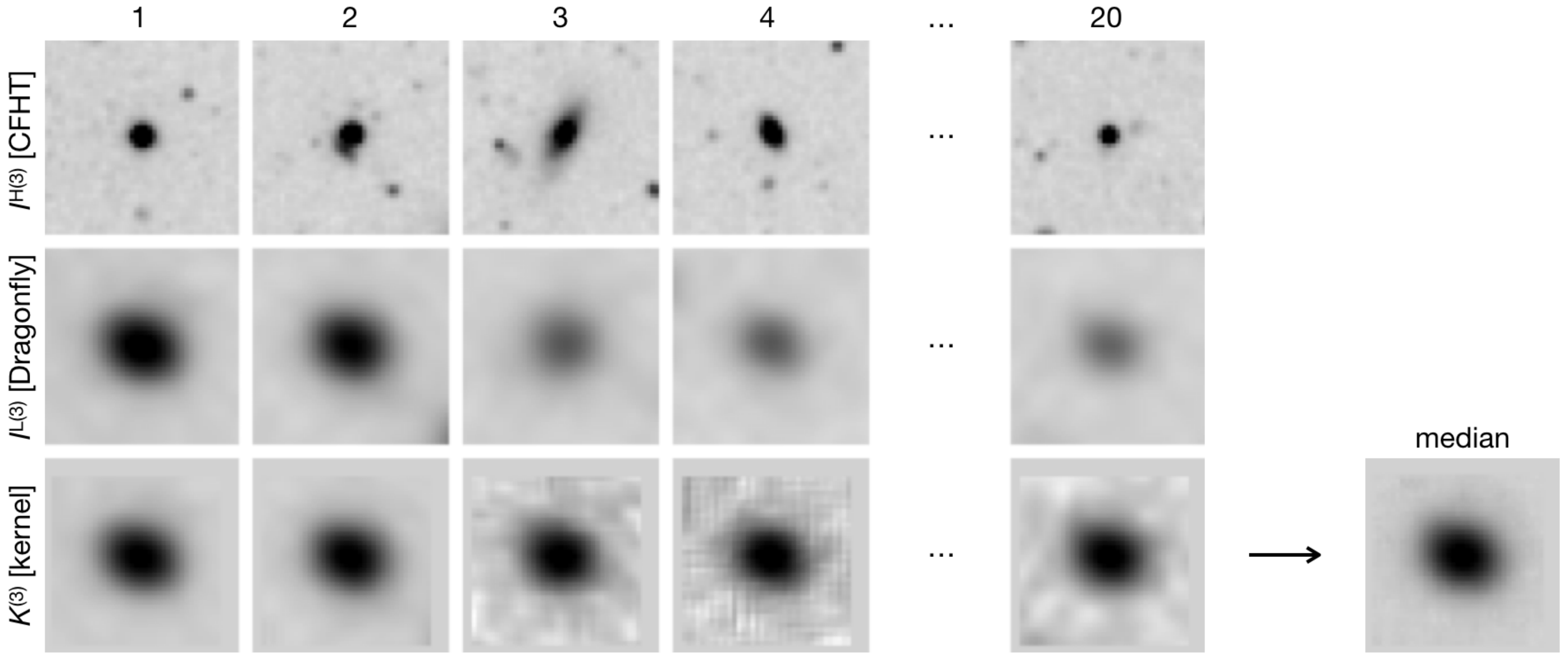}
  \end{center}
\vspace{-0.2cm}
    \caption{
Creation of the convolution kernel $K$ for the M101 image that is used as a demonstration
of the implementation. {\em Top row:} Bright, non-saturated, compact objects
in the high resolution CFHT image. Most of the 20 objects are stars. The third
and fourth example objects were chosen to illustrate that the few compact galaxies produce very similar
kernels as stars, as expected.
{\em Middle row:} Dragonfly images of the same objects. {\em Bottom row:}
Convolution kernels to go from the top row to the middle row,
derived by an inverse Fourier transform of the quotient of the Fourier transforms
of the two images. The convolution
kernel that is used in the analysis is the median of
these twenty individual kernels.
Each panel is $29\arcsec \times 29\arcsec$, sampled at $0\farcs 67$\,pix$^{-1}$
(1/3 of the pixel scale of the Dragonfly M101 data).
}
\label{kernel.fig}
\end{figure*}

\subsection{Convolution kernel}

The convolution kernel to bring the high resolution model to the Dragonfly
resolution is created using the Fourier quotient method, as explained
in \S\,\ref{mrf_method.sec}. The actual kernel that is used is the median of
a large number (typically 20--30) of individual kernels. These are created
from image segments that are centered on bright, unsaturated objects and are spread
evenly over the image.

In practice the following steps are taken. First, all objects in the $I^{\rm H(3)}$
catalog that are close to the edge of the image or have a flux greater
than a user-defined value are discarded.
This
step serves to remove bright objects that are saturated in either $I^{\rm H(3)}$
or $I^{\rm L(3)}$ or for other reasons should not be taken into account.
Typical values are $0.01<f_{\rm max}<0.1$, with $f_{\rm max}$ the user-defined brightness with
respect to the fifth-brightest object in the catalog. The optimal
value depends on the depth of the images, the
seeing, the number of very bright stars in the field, and the minimum S/N ratio that
is required: the parameter $f_{\rm max}$ effectively controls from which part of the
luminosity function the kernel objects are drawn. 
An appropriate choice of $f_{\text{max}}$ is critical for obtaining
a clean residual map $R$.

The next-brightest $N_{\rm kernel}$ objects in the catalog are selected, with
$N_{\rm kernel}$ a user-defined parameter. Several of these objects will be large
galaxies or close
to other bright stars or galaxies, making them less suitable as inputs to the
median kernel. We remove such objects by requiring that $b/a>0.6$ and $0.8<
I^{\rm H}_{i, \rm section}/I^{\rm H}_{i, \rm catalog}<1.5$, that is, that they are fairly round and that the flux
in the image section
that is used for the Fourier transforms is close to the catalog flux.
For the remaining objects (nearly all of them isolated stars)
we determine the median kernel using Eqs.\ \ref{fft1.eq}
and \ref{fft2.eq}. We note
that the individual kernels, and
the median kernel, are {\em not} normalized to unity. Instead,
the integrated flux is
equal to the flux ratio (in ADU) between the high resolution and low resolution
images. The kernel therefore implicitly accounts for any relative errors in the zeropoints
of the two images.
The process is illustrated in Fig.\ \ref{kernel.fig} for
the M101 example. In this example, 17 of the 20 compact objects are stars; as shown
in columns 3 and 4 of Fig.\ \ref{kernel.fig} the few compact galaxies that satisfy
the criteria produce identical kernels
as stars (as expected from the Fourier quotient method).

\begin{figure*}[htbp]
  \begin{center}
  \includegraphics[width=1.0\linewidth]{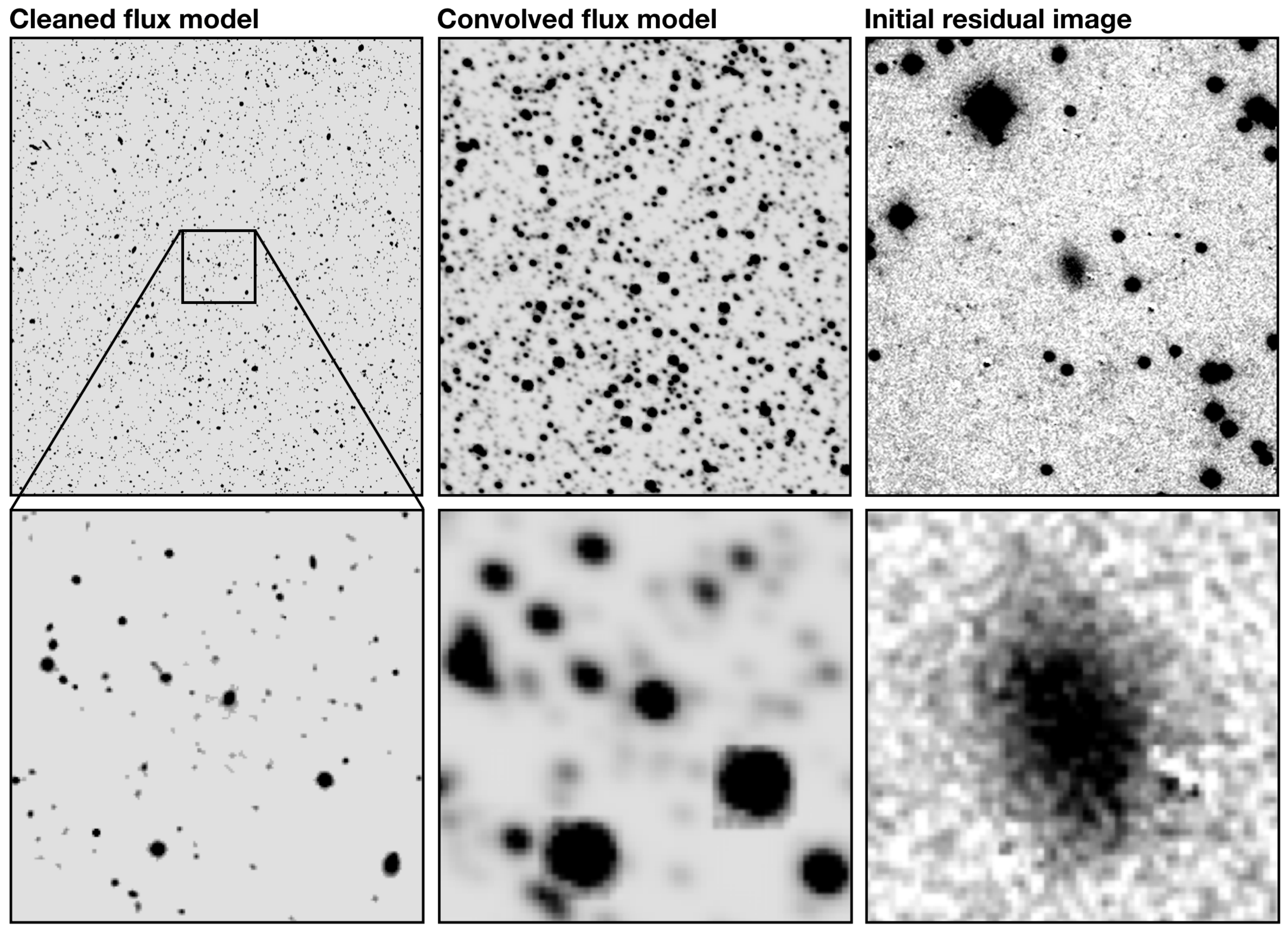}
  \end{center}
\vspace{-0.2cm}
    \caption{
{\em Top left:} Flux model $F^{\rm H(3)}$ after removing bright stars and
low surface brightness emission on the scale of the Dragonfly PSF.
{\em Top middle:} Flux model convolved with kernel $F^{\rm L}$, resampled
to the Dragonfly pixel scale. {\em Top right:} Residual image $I^{\rm L}-
F^{\rm L}$. Most compact sources are removed from the Dragonfly data, while
bright stars and low surface brightness emission are preserved. The
bright stars are treated
separately. {\em Bottom row: zoomed views.}
}
\label{df3_model.fig}
\end{figure*}

\subsection{Removal of low surface brightness emission from the model}

Prior to convolving the flux model with the kernel two classes of sources
are removed from it: bright stars (see \S\,\ref{star_removal.sec}) and
objects that have the spatial
scale of the Dragonfly PSF and are fainter than a particular
surface brightness threshold. This step serves to ensure that {\em only} compact
objects are subtracted from the Dragonfly data. Without it, object catalogs created
from filtered Dragonfly data would be difficult to interpret as any low
surface brightness objects that were entirely or partially detected in the
high resolution images would no longer be present in the images.
The M101-DF3 example is a good illustration: as can be seen in the bottom
right panel of Fig.\ \ref{df3_intro.fig} small sections of the galaxy are
in the initial flux model, as they are  identified  as a clump of 
low surface brightness objects with
SExtractor.

We remove low surface brightness objects in the following way. An image is created that is
sensitive to the spatial extent of objects with respect to the Dragonfly PSF:
\begin{equation}
E = \frac{F^{\rm H(3)}}{F^{\rm H(3)} \ast K}.
\end{equation}
If $\langle E\rangle \sim 1$, with the angled brackets indicating the average
over all pixels belonging to the object according to the segmentation map, its
spatial extent is larger than the Dragonfly PSF. If $\langle E\rangle \gg 1$
it is a compact object that should be retained in the model and subtracted
from the Dragonfly data.

Objects are removed from the model (and hence retained in the 
Dragonfly data) if they satisfy the following two criteria: $\langle E\rangle <
E_{\rm lim}$, with $E_{\rm lim}$ a user-defined value that is set to $E_{\rm lim}=6$
for the M101 data; and $\langle F^{\rm H(3)}\rangle < F_{\rm lim}$, with
$F_{\rm lim}$ a maximum surface brightness. 
In the case of the M101 example
this limit was set to the equivalent of a surface brightness of
$\mu_r = 24.0$\,\ma. The results are very similar when changing these limits
by factors of $\sim 2$. Optionally a minimum object area can be specified as an
additional criterion.

\begin{figure*}[htbp]
  \begin{center}
  \includegraphics[width=1.0\linewidth]{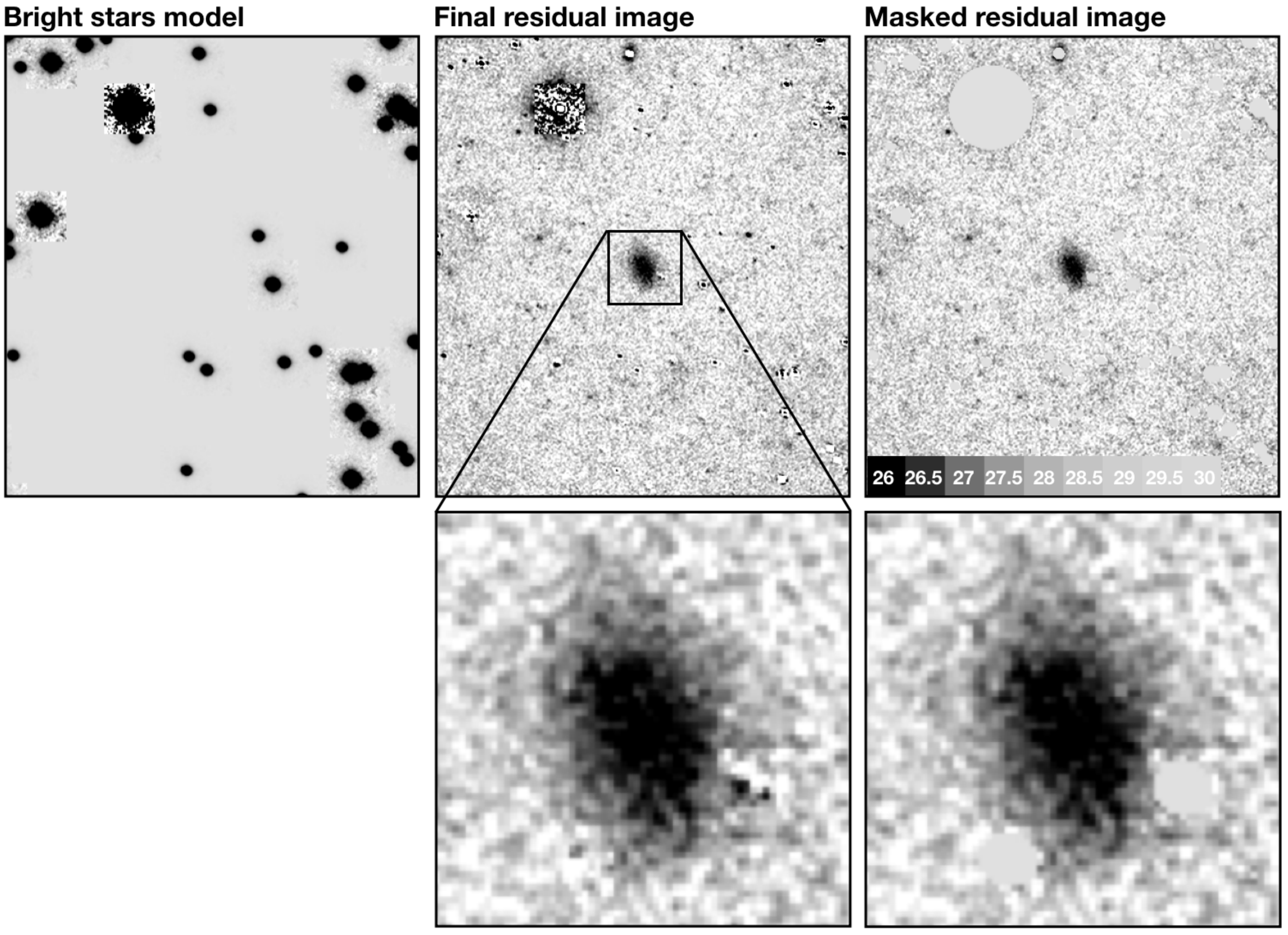}
  \end{center}
\vspace{-0.2cm}
    \caption{
{\em Top left:} Model for the bright stars that were removed from the
flux model.
{\em Top middle:} Residual image after subtracting the bright
stars. {\em Top right:} Residual image after masking the regions
corresponding to the brightest pixels
in the convolved flux model, as well as stars with $m_r<12.5$ (only one
in the region that is shown). The bar at the bottom indicates the conversion
of grey level to $r$-band surface brightness, in units of mag\,arcsec$^{-2}$.
{\em Bottom row: zoomed views.}
}
\label{df3_final.fig}
\end{figure*}

\subsection{Removal of bright stars from the model}
\label{star_removal.sec}

The MRF procedure is very effective in removing faint compact sources but it
is not suitable for the removal of very bright stars. There are two reasons for
this. First, the kernel is only $36\arcsec \times 36\arcsec$, and this is smaller than the
extent of the wings of bright stars. Creating larger kernels is possible (the
kernel size is a user-defined parameter in the {\tt mrf} code) but the S/N
ratio in the outer parts is low for the unsaturated stars and
galaxies that are used to create it.
The second issue is saturation, particularly
in the high resolution data. Typical high resolution images (including the CFHT
data used in the example) are relatively long exposures on large telescopes, and many
of the brighter stars and galaxies are satured in their centers. This leads to several
distinct problems. The first is bleeding in the high resolution data, as is happening
for the bright star in the upper left of the example image in Fig.\ \ref{df3_intro.fig}.
These erroneous features make their way into the residual image as strongly negative
pixels as they are part of the flux model. The second is that the flux of these
objects in the high resolution data is lower than the true brightness, sometimes
severely so. As a result, bright stars in the flux model are too faint, leaving
positive residuals in the final star-subtracted image.
The third is centroiding errors, which cause strong positive and negative residuals.

For these reasons 
the {\tt mrf} code removes the brightest objects from the flux
model, down to a user-defined magnitude.\footnote{This magnitude is not necessarily equal to
the true brightness of the stars; it is the SExtractor AUTO magnitude as measured from
the high resolution image, with no
attempt to correct for saturation.} The code outputs a list of the objects
(mostly stars) that are excluded from the model. At a later stage
they are identified
in, and removed from, the Dragonfly image (\S\,\ref{halosub.sec}).\footnote{Some very bright objects
can be galaxies, and objects of interest. If desired, these can be manually specified by their RA and DEC and retained,
as explained in \S\,3.2.}
In the M101-DF3 example the magnitude limit that is used is $m_r<17.5$.
The ``cleaned'' flux model, with low surface brightness emission and bright stars removed,
is shown in the left panels
of Fig.\ \ref{df3_model.fig}.

\subsection{Convolution of the model and subtraction}

The cleaned flux model is convolved with the median kernel to match the Dragonfly
resolution,
\begin{equation}
F^{\rm L(3)} = F^{\rm H(3)} \ast K,
\end{equation}
and subtracted:
\begin{equation}
R^{\rm (3)} = I^{\rm L(3)} - F^{\rm L(3)}.
\end{equation}
The residual is then binned to the original grid of the Dragonfly data to create $R$.
The convolved model $F^{\rm L}$ (rebinned to the Dragonfly resolution) is shown
in the middle panels of Fig.\ \ref{df3_model.fig}. The residual $R$ is
shown in the right panels. The residual image consists of low surface brightness
emission, the bright stars that were purposely left in the image,
and artifacts due to imperfect modeling of the centers of subtracted stars.

\subsection{Subtraction of bright stars from the residual image}
\label{halosub.sec}

Bright stars are modeled and subtracted in a way that is analogous to that described
in {van Dokkum} {et~al.} (2014), {Merritt} {et~al.} (2014), and {Merritt} {et~al.} (2016a). First, a catalog of
objects in the residual image is
created with SExtractor. Next, the catalog is cross-matched with the objects that were
removed from the flux model, and only those that are matched are retained. This matching
requirement ensures
that no spurious bright objects in the high resolution image are subtracted from
the Dragonfly data, and that objects that were already subtracted as part
of the convolved flux model are not subracted twice.  A model PSF is created by
taking the median of image cutouts of the matched stars, after shifting
them to the same sub-pixel position and normalizing them. The normalization is
not done by the total (or AUTO) flux but by the flux in an annulus between
$r=3$\,pix ($6\arcsec$) and $r=6$\,pix ($12\arcsec$), to avoid any saturated
pixels. In Dragonfly images even very bright stars are typically not saturated
beyond $r=6\arcsec$, due to the steep fall-off of the PSF (see {Abraham} \& {van Dokkum} 2014).

For each bright object the model PSF is scaled to its total flux (taking saturation
into account)
and placed at its location, with sub-pixel accuracy.  The size of the model PSF
is a user-defined parameter; in this example we use $48\times 48$
pixels, or $96\arcsec \times 96\arcsec$. The image containing the scaled models
is shown in the top left panel of Fig.\ \ref{df3_final.fig}. This image is
subtracted to create the final residual image $R$, shown in the middle
panels of Fig.\ \ref{df3_final.fig}.

Although the subtraction is adequate for this M101 region, some Dragonfly data include
individual very bright stars whose light at large radii cannot be modeled with this
procedure. The {\tt mrf} code includes the option to extend the stacked 2D model PSF
to radii $>30\arcsec$ (commonly referred to as the aureole; see King 1971)
using an analytic function. As discussed in Q.~Liu et al., in preparation,
the form of the aureole function is a composite
of powerlaws:
\begin{equation}
\label{aureole.eq}
I_{\rm au}(R)= \sum_{i=0}^{k}{\frac{A_i}{R_i^{-n_i}}R^{-n_i}},
\end{equation}
with $k=2$. The constants $A_i$ are not fit parameters but fully determined
by the values of $R_i$, $n_i$, and the total magnitude of the star.
The transition radii $R_i$ and powerlaw slopes $n_i$ are
adjustable in the code as they depend on the atmospheric conditions at the
time of the observation.
In \S\,4.3 we show an example of a previously-published image with several bright stars, where the
analytic function improves the subtraction.
In the M101 image discussed here the analytic function extension was not used.

\begin{figure}[htbp]
  \begin{center}
  \includegraphics[width=0.96\linewidth]{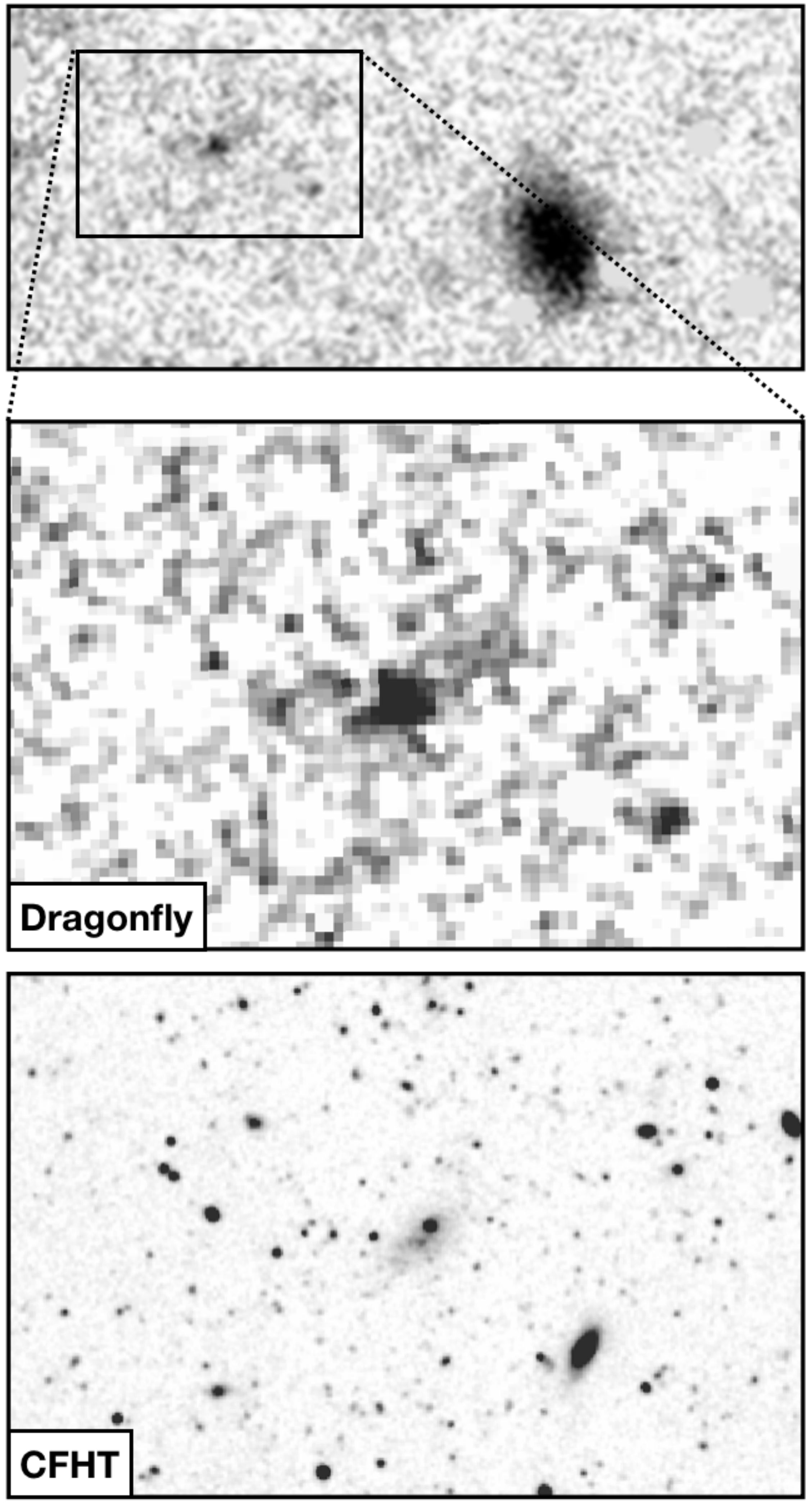}
  \end{center}
\vspace{-0.2cm}
    \caption{
A possible very faint nucleated dwarf galaxy in the model-subtracted M101-DF3 image.
The galaxy has an absolute magnitude
of $M_r \approx -8.0$ if it is at the distance of M101. 
Note that the very bright and large object in the top panel is
M101-DF3 itself, which was unknown prior to 2014 and discovered in
these Dragonfly data (Merritt et al.\ 2014).
}
\label{new_dwarf.fig}
\end{figure}

\subsection{Masking of artifacts in the final residual image}
\label{mask.sec}
The final residual image is not free of artifacts but their locations can be
robustly predicted. They mostly occur in locations were bright stars were
subtracted, as even a small error in the flux model can produce a large
residual in regions where the counts are high.\footnote{We find that
errors are typically $\lesssim 5$\,\% per pixel.}
All regions in the residual image are masked where the subtracted
model $F^{\rm L} > F_{\rm lim}$, with $F_{\rm lim}$ a user-defined parameter.
Only the model flux is used to determine whether to mask pixels
in the residual image, and not the residual itself;
this is
to prevent the masking of relatively bright
objects that may be present after the subtraction of the model.
The mask can optionally be extended to include pixels neighboring those
that are brighter than
$F_{\rm lim}$.

The very brightest stars are also masked, using a circular mask with a user-defined
radius for objects down to a user-defined magnitude. For the M101-DF3 example
we use a limit of $m_r<12.5$ and a radius of 40\,pixels ($80\arcsec$). Only one
object, the $m_V=10.9$ star TYC 3852-845-1, falls in this category.
The masked residual image of M101-DF3 is shown in the right panels of
Fig.\ \ref{df3_final.fig}.
We note that the required masking impacts far less pixels than would be the case if
the convolved flux model had not been subtracted. In the M101-DF3 example,
a simple mask applied to the original Dragonfly image would cover $\sim 20\times$ more
pixels than the mask shown in Fig.\ \ref{df3_final.fig}. For a visual impression, compare
the left panels of Fig.\ 2 to the right panels of Fig.\ 4.

The final masked image shows M101-DF3 itself very clearly. In addition, several other
faint sources are still present after subtraction of the convolved flux model and
masking. These fall in several categories: compact variable sources, such as variable
stars and active galactic nuclei; low surface brightness emission associated with galaxies,
such as tidal tails and stellar halos; and low surface brightness dwarf galaxies.
One of the brightest examples is shown in Fig.\ \ref{new_dwarf.fig}. This low surface
brightness object
is a possible nucleated dwarf galaxy, at  ${\rm RA}=14^{\rm h}3^{\rm m}27\fs38$,
${\rm DEC}=53\arcdeg{}37\arcmin{}51\farcs8$ (J2000). It is not in the sample
of faint low surface brightness objects of Carlsten et al.\ (2019), and given its
proximity to M101 it is unlikely that it is a member of the background NGC\,5485 group
(see Merritt et al.\ 2016b, Karachentsev \& Makarova 2019).
Its $r$-band magnitude in the
Dragonfly data is $m_r\approx 21.2$.\footnote{Note that the star that is visible in the CFHT image in the
Northwest quadrant of the galaxy was subtracted in the MRF process.}
If it is at the distance of M101 its absolute
magnitude is $M_r\approx -8.0$, which would place it among the lowest luminosity
dwarf galaxies yet identified
outside of the Local Group (see, e.g., Smercina et al.\ 2017, Mihos et al.\ 2018).

\begin{figure*}[htbp]
  \begin{center}
  \includegraphics[width=1.0\linewidth]{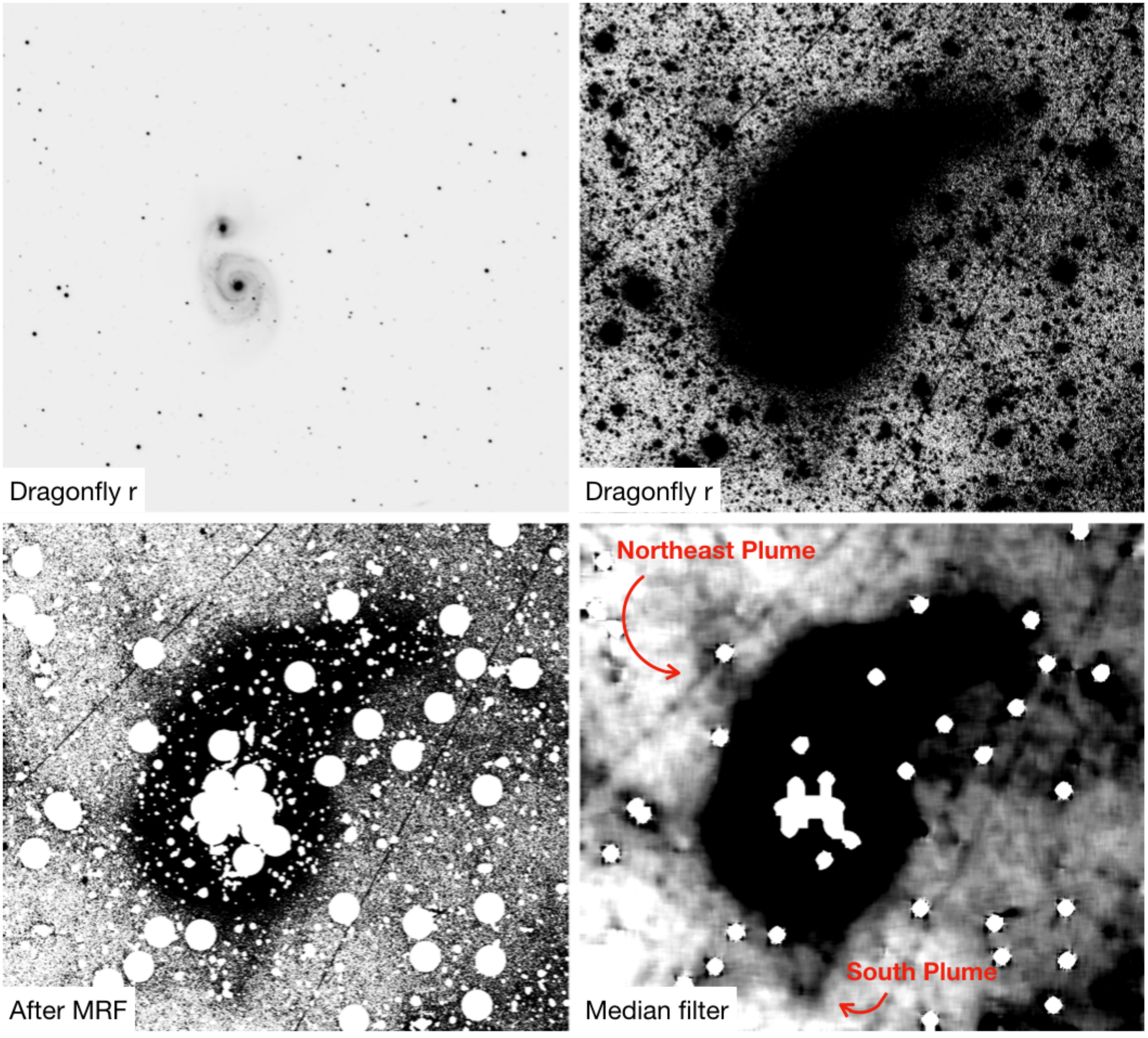}
  \end{center}
\vspace{-0.2cm}
    \caption{
{\em Top panels:} Dragonfly $r$-band image of M51, with two different scalings.
The displayed image is a $42\arcmin\times 38\arcmin$ cutout.
{\em Bottom left:} After multi-resolution filtering. {\em Bottom right:} Median
filtered version of the multi-resolution filtered image. Two 
$\mu_B\sim 29$\,\ma\ features
that were discovered by {Watkins}, {Mihos}, \&  {Harding} (2015) are indicated.
}
\label{m51.fig}
\end{figure*}

\begin{figure*}[htbp]
  \begin{center}
  \includegraphics[width=1.0\linewidth]{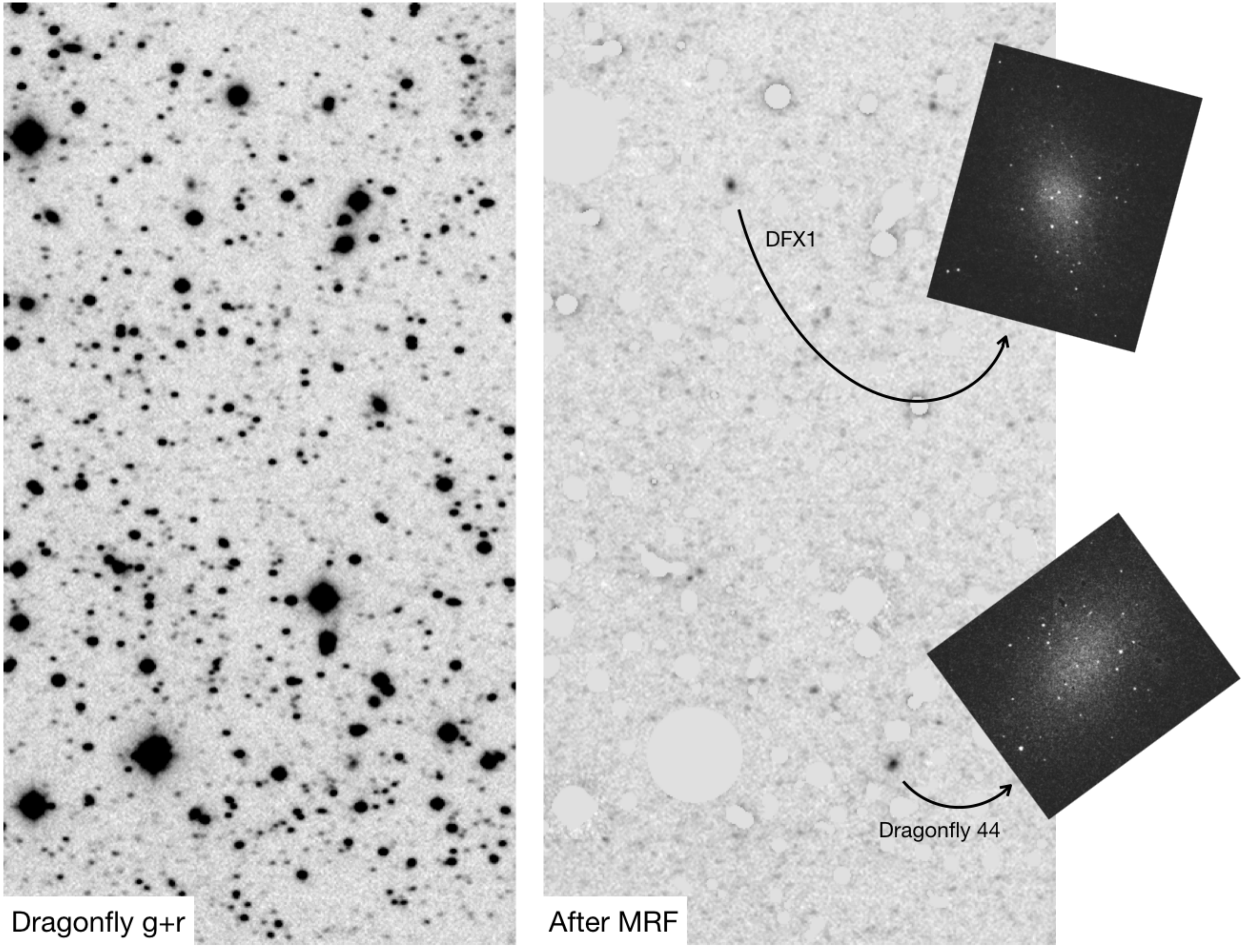}
  \end{center}
\vspace{-0.2cm}
    \caption{
{\em Left:} Small ($12\farcm 4 \times 22\farcm 9$)
section of the Dragonfly $g+r$-band Coma image 
that was used
in {van Dokkum} {et~al.} (2015) to identify 47 ultra-diffuse galaxies (UDGs) in the
cluster. This section contains two UDGs, Dragonfly 44 and DFX1.
{\em Right:} The same region after multi-resolution filtering, using 300\,s
CFHT images to construct the flux model. The two UDGs are now the brightest
objects in the image. The insets show their HST images, with extended
sources masked (see {van Dokkum} {et~al.} 2017).
}
\label{coma.fig}
\end{figure*}

\begin{figure*}[htbp]
  \begin{center}
  \includegraphics[width=0.6\linewidth]{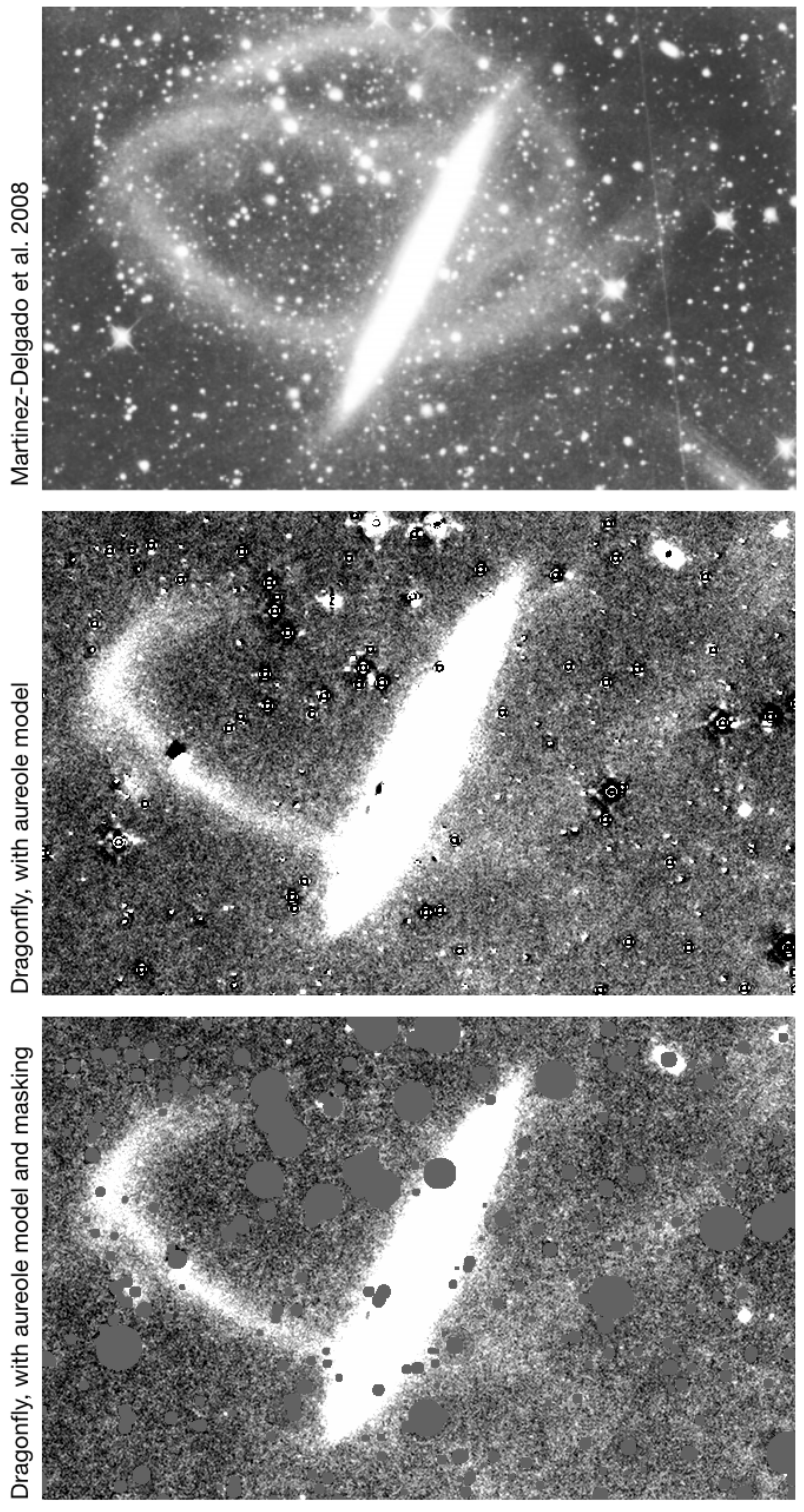}
  \end{center}
\vspace{-0.2cm}
    \caption{
{\em Top:} The stellar stream system of NGC\,5907, as imaged
by Mart{\'{\i}}nez-Delgado et al.\ (2008).
{\em Middle:} Dragonfly $g+r$ image of the same region, after subtracting compact emission.
The {\tt mrf} code was run with aureole modeling.
{\em Bottom:} The same as the middle image but with areas masked where the flux model
exceeds a certain threshold. Thanks to the
aureole modeling this image shows less residuals around bright stars than
that presented in van Dokkum et al.\ (2019).
}
\label{n5907.fig}
\end{figure*}

\section{Other Examples}

Here we apply the MRF technique to three additional Dragonfly datasets, to demonstrate
the method in other contexts.  The high resolution images of the three objects that are
used in the analysis are
shown in Fig.\ \ref{highres.fig} in Appendix B. 
Other applications of the MRF technique on Dragonfly observations of nearby
galaxies may be found in van Dokkum et al.\  (2019), Gilhuly et al.\ (2019), and
J.~Li et al., in preparation.

\subsection{M51}

The interacting galaxy pair NGC\,5194 and NGC\,5195 (M51) was our first-light target
with the upgraded 48-lens Dragonfly telescope and was observed on July 3 2016.
Not all the lenses were operational all the time. The 
equivalent exposure
time for a 100\,\% operational array with 24 lenses in each filter is 6950\,s
in $g$ and 6000\,s in $r$. A section of the $r$-band image, centered on M51, is shown
in the top panels of Fig.\ \ref{m51.fig}, with two different
scalings. The reduction was performed with an
early version of our pipeline and did not remove satellite trails.

The pair is embedded in an extensive tidal debris field. Despite many years
of observations of this system, {Watkins} {et~al.} (2015) discovered
two previously-unknown tidal features in extremely deep images obtained with the
Burrell-Schmidt telescope on Kitt Peak: the ``South Plume'' and the ``Northeast Plume'',
with surface brightness $\mu_B\sim 29$\,\ma\ 
(see {Watkins} {et~al.} 2015).
Here we determine whether these features can be seen in the Dragonfly data as well,
after applying the MRF technique.

The {\tt mrf} code is run with very similar parameters as in the M101-DF3 example.
The high resolution data is again from CFHT: M51 was observed for 525\,s 
in the $r$-band with Megacam 
on April 4 2007. We obtained the reduced image from the CFHT Archive.
The results are shown in the bottom left panel of Fig.\ \ref{m51.fig}. After removal
of the compact sources tidal features are more easily detected. This is illustrated
in the bottom right panel, which shows the MRF image after applying a $21\times 21$
pixel ($52\farcs 5 \times 52\farcs 5$) median filter. The two plumes discovered
by {Watkins} {et~al.} (2015) are readily seen in the Dragonfly image (see
also Rich et al.\ 2019).

\subsection{Ultra-diffuse galaxies in the Coma cluster}

In 2015 a large population of faint, intrinsically-large galaxies was identified in the
Coma cluster with the Dragonfly telescope
({van Dokkum} {et~al.} 2015). These objects have sizes that are close to the
Dragonfly PSF ($\sim 5\arcsec$) and it was
difficult to isolate them against a background of many thousands of other
objects. Later studies used wide-field imaging data from conventional
telescopes with much better seeing to find UDGs in clusters and the general
field, with great success (see, e.g., {Koda} {et~al.} 2015; {van der Burg} {et~al.} 2017; {Rom{\'a}n} \& {Trujillo} 2017; {Danieli} \& {van Dokkum} 2019).
Here we return to the original Dragonfly Coma data, to determine whether the MRF
technique is able to efficiently identify these objects.

We use CFHT $g$ and $i$ images for creating the flux model. These
data are described in {van Dokkum} {et~al.} (2015) and
{Danieli} \& {van Dokkum} (2019). Neither of these filters is a good match to the Dragonfly
ones, and instead we derive a color term to go from CFHT $g$ and $i$ to the
sum of the Dragonfly $g+r$ images. Nevertheless, we find that the quality
of the image subtraction
is limited by the mismatch between the filters.

The Dragonfly $g+r$ image is shown in the left panel of Fig.\ \ref{coma.fig}.
We focus on a $12\farcm 4 \times 22\farcm 9$ section that contains two
UDGs, Dragonfly 44 and DFX1 (see {van Dokkum} {et~al.} 2017). The galaxies are not
easily distinguished from other sources in the field, which is why
{van Dokkum} {et~al.} (2015) used a complex multi-stage process to eliminate faint objects
from an initial candidate list of 6624.
The right panel shows
the same region after applying the {\tt mrf} code, with default parameters.
The two UDGs are now the two brightest objects in this region, and they can
readily be identified using standard image detection software. 

\subsection{The tidal stream of NGC\,5907}

In a previous paper (van Dokkum et al.\ 2019) we applied 
the {\tt mrf} methodology to a deep Dragonfly image of the edge-on galaxy NGC\,5907.
The galaxy has a curved tidal stream, discovered by Shang et al.\ (1998).
Mart{\'{\i}}nez-Delgado et al.\ (2008) found that the stream consists not of one
but of two loops, wrapping twice around
the galaxy. However, the Dragonfly data, as well as recent
observations by Muller et al.\ (2020), failed to confirm this second loop.

There are several bright stars in the NGC\,5907 field, and some of these are close
to the purported second loop (see van Dokkum et al.\ 2019). In our published
{\tt mrf} analysis no attempt was made to subtract the extended aureoles of these stars.
Since then we implemented the optional analytic PSF extension of Eq.\ \ref{aureole.eq},
and here we show the improved subtraction that results from this. We use transition
radii of $R_0=5\arcsec$, $R_1=71\arcsec$, and $R_2=151\arcsec$, and powerlaw indices
of $n_0=3.24$, $n_1=2.53$, and $n_2=1.22$. These values are determined from
a group of very bright stars in the wider Dragonfly field,
$\approx 45\arcmin$ to the Southwest of NGC\,5907 itself (see Fig.\ 1 of van Dokkum
et al.\ 2019). The powerlaw model is tied
to the empirical stacked 2D PSF at $R=30\arcsec$, and truncated at
$R=20\arcmin$ to ensure that the integrated model flux is finite.\footnote{The outer
powerlaw index $n_2$ is relatively shallow compared to other Dragonfly fields, indicating
that there was some cirrus at the time of observation and/or that the lenses were somewhat dusty.
See Q.~Liu et al., in preparation, for details.}

The results are shown in Fig.\ \ref{n5907.fig}. Here we focus on the small region of
the Dragonfly image that overlaps with the image of Mart{\'{\i}}nez-Delgado et al.\ (2008);
for a complete analysis we refer the reader to van Dokkum et al.\ (2019).
The aureole modeling improves the subtraction of bright stars compared to the
version of the code that was used in our previous analysis.

\section{Summary}

We present a straightforward method to use high resolution data to remove
compact sources of emission from low resolution data. The method reliably distinguishes
blended compact sources from low surface brightness emission, something that is
very difficult to achieve with standard techniques applied to low resolution
images (see, e.g., van Dokkum et al.\ 2015).
It is implemented in the
{\tt mrf} Python package,
which we make publicly available.\footnote{\url{https://github.com/AstroJacobLi/mrf}}
In Appendix A we discuss two variations of the technique: self-MRF, where a
smoothed version of an image takes the place of the low
resolution image $I^{\rm L}$, and cross-MRF,
where two high resolution images are available from different telescopes and
one is smoothed to create $I^{\rm L}$. Both are included in the public
distribution of the {\tt mrf} code.

The {\tt mrf} tool 
is important for the correct interpretation of data from the Dragonfly
Telephoto Array, as crowding is a significant problem for this telescope.
By removing the confusing signal of blurred compact objects the benefits of
Dragonfly, such as its wide field of view, low false positive rate, and its
excellent control of scattered light, can be fully utilized. 
High resolution images of sufficient depth are generally readily available, as
Dragonfly's sensitivity to compact objects is equivalent to that of a 1\,m telescope
with $\approx 5\arcsec$ seeing.
MRF has
been applied to the analysis of data from the Dragonfly Edge-on
Galaxy Survey (van Dokkum et al.\ 2019; Gilhuly et al.\ 2019), and
is an integral
part of the Dragonfly Wide Field Survey (Danieli et al.\ 2020),
which aims to identify faint dwarf galaxies in a 400\,degree$^2$ area.
Looking ahead,
images with a resolution of $\sim 0\farcs 2$ will be routinely available over large
parts of the sky from EUCLID and/or WFIRST, and those data may serve
as the high resolution flux model to aid searches for faint extended sources
in deep images from large telescopes on the ground.

The present paper does not discuss systematic uncertainties
which might be introduced by the technique. Filtered frames should be treated with
caution: photometry at the location of removed
objects is more uncertain than photometry in ``pristine'' regions, and incomplete
removal of compact objects could lead to false positives in low surface brightness
galaxy searches. How severe these issues are depends on the specific dataset and
the specific science questions that are asked. We urge users of filtered data
to incorporate 
the flux model and residual mask created by {\tt mrf} in their
object detection and photometry codes. In T.~Miller et al., in preparation,
we assess the recovery of total fluxes of realistic galaxy images injected
in data from the Dragonfly Wide Field Survey. Performing photometry on the
MRF-cleaned images yields total fluxes with no systematic bias and a random
scatter of $0.08$\,magnitudes.


\acknowledgements{
We thank the referee for a thorough report that improved the manuscript. Support from NSF
grant AST-1613582 is gratefully acknowledged.
The {\texttt{mrf}} code makes use of \href{http://www.numpy.org}{\code{NumPy}}, a package for scientific computing with Python (Walt, Colbert, \& Varoquaux 2011); \href{https://www.scipy.org}{\code{SciPy}}, an open source scientific library ({Virtanen} {et~al.} 2019); \href{https://matplotlib.org}{\code{Matplotlib}}, a 2D plotting library ({Hunter} 2007); \href{http://www.astropy.org}{\code{Astropy}}, a community-developed core Python package for Astronomy ({Astropy Collaboration} {et~al.} 2018); \href{https://sep.readthedocs.io/en/v1.0.x/}{\code{sep}}, a Python library for Source Extraction and Photometry ({Bertin} \& {Arnouts} 1996; Barbary 2016); \href{http://galsim-developers.github.io/GalSim/_build/html/index.html}{\code{GalSim}}, a galaxy image simulation toolkit;
and \href{https://iraf-community.github.io/}{\code{IRAF}}, the Image Reduction and Analysis Facility ({Tody} 1986, 1993).
}

\begin{appendix}

\section{Variations: self-MRF and cross-MRF}

The MRF code was developed to isolate low surface brightness emission by removing
all compact sources from the image. The specific application, discussed extensively
in the main text, is the ``cleaning" of Dragonfly images using seeing-limited wide-field
ground based data. However, in many cases no such intrinsically-low resolution imaging will
be available, and most aspects of the MRF methodology for isolating low surface brightness emission can still be applied in those instances. In self-MRF, a smoothed version of an image is used as the low resolution model. This smoothed image takes the place of $I^{\rm L}$, and the code runs in the same way as in its standard implementation. The smoothing kernel can be tuned to the particular structures that the user intends to isolate; by experimenting on known dwarf galaxies in Hyper Suprime-Cam data
we find that a convolution with an exponential kernel provides the best results. We note that self-MRF is essentially a variation of standard low
pass filtering approaches (see, e.g., Greco et al.\ 2018b).

Another common situation is where two high resolution images are available, from different
telescopes. Many extragalactic surveys cover overlapping sky areas, such as the GAMA
fields (Driver et al.\ 2011). In cases where, for instance, data from
Subaru and CFHT are available one data set can be smoothed to form $I^{\rm L}$ and
the other can assume the role of $I^{\rm H}$. In this application, dubbed
cross-MRF, the survey with
the best low surface brightness sensitivity (typically, the one with the best sky
subtraction) can be used for $I^{\rm L}$. An important advantage over self-MRF is
that artifacts (such as diffraction spikes) are usually not present at the same location in two independent observations. Demonstrations of both self-MRF and cross-MRF are provided with the \texttt{mrf} GitHub distribution.

\section{High resolution images of the three examples}

\begin{figure*}[htbp]
  \begin{center}
  \includegraphics[width=0.85\linewidth]{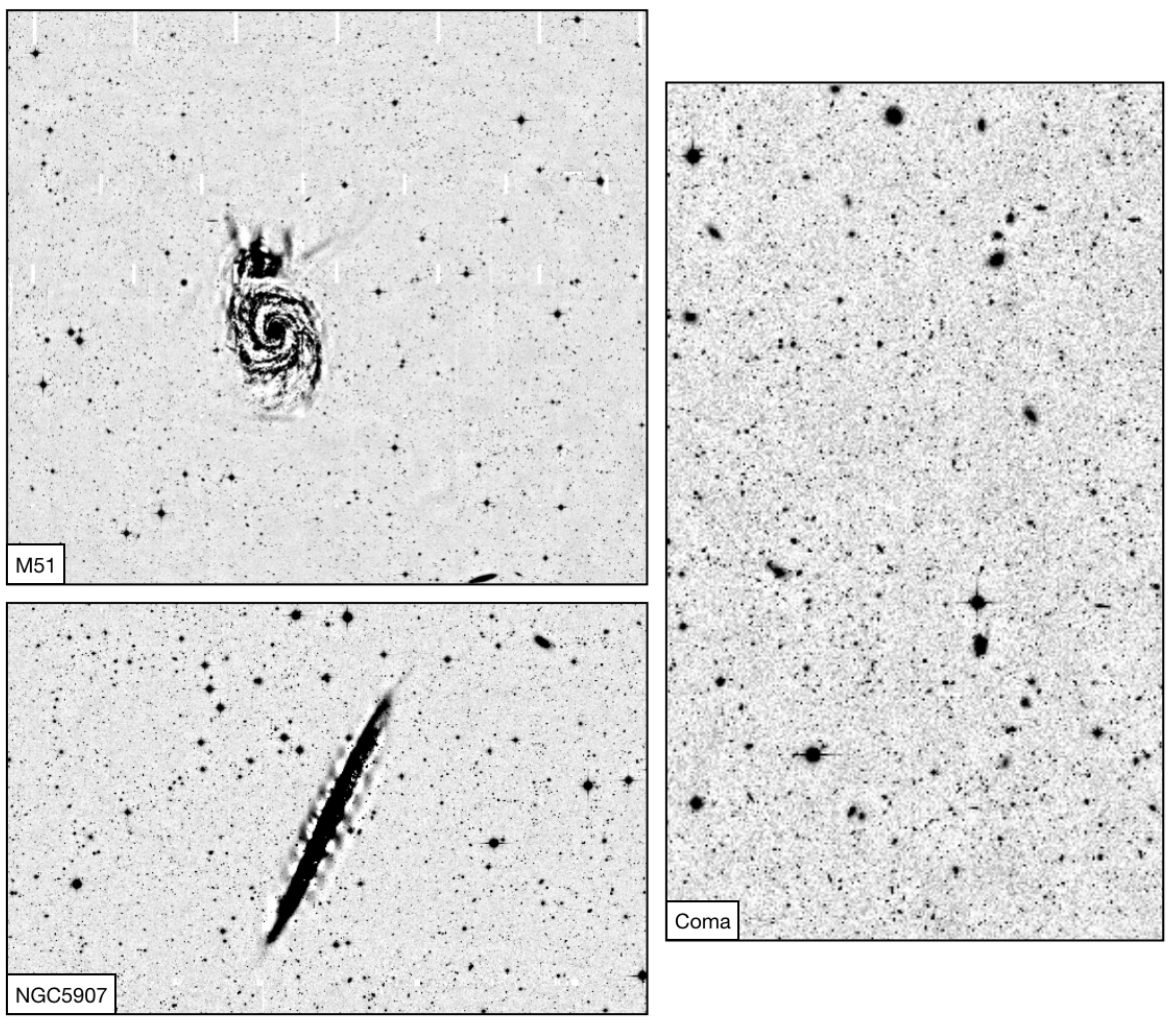}
  \end{center}
\vspace{-0.2cm}
    \caption{
The high resolution images that are used for the multi-resolution filtering of the Dragonfly images in Figs.\ 7, 8, and 9. Note the complementarity of these CFHT images and the Dragonfly images: the CFHT data have excellent point source sensitivity, but all spatially-extended low surface brightness emission was removed in the reduction process.
}
\label{highres.fig}
\end{figure*}

In Fig.\ \ref{highres.fig} we show the public CFHT images of the three fields that are analyzed
in \S\,4. These CFHT images are the high resolution data that are used to create the flux models.
It is striking that no large scale low surface brightness emission is visible around M51 and
NGC\,5907. This is caused by the data reduction procedures that are applied to these publicly available CFHT data; as
described in Gwyn (2008), the MegaPipe pipeline subtracts spatially-extended emission. 
For our purposes this oversubtraction is not an issue, but 
we note here that far superior results can be obtained from CFHT with Elixir-LSB, a low surface brightness-optimized reduction process (J.~C.\ Cuillandre et al., in preparation). 

\end{appendix}


\end{document}